\documentstyle[prb,aps,twocolumn]{revtex} 
\begin{document} 
\draft 
\flushbottom 
 
%**************************************************** 
\flushbottom 
\twocolumn[\hsize\textwidth\columnwidth\hsize\csname@twocolumnfalse\endcsname 
%**************************************************** 
 
%\bibliographystyle{plain} 

\title{Transition from Diffusive to Localized Regimes in Surface 
Corrugated Optical Waveguides.} 

\author{A.  Garc\'{\i}a-Mart\'{\i}n, J.A.  Torres, J.J.  
S\'{a}enz\cite{byline}} 

\address {Departamento de F\'{\i}sica de la Materia Condensada\\ and 
Instituto de Ciencia de Materiales ``Nicol\'{a}s Cabrera''.\\ 
Universidad Aut\'{o}noma de Madrid,  E-28049 Madrid, Spain.\\ } 
\author{ M. Nieto-Vesperinas}

\address { Instituto de Ciencia de Materiales. 
Consejo Superior de Investigaciones Cient\'{\i}ficas.\\ Campus de Cantoblanco,  
E-28049 Madrid, Spain. }
 
\date {\today} 
\maketitle 

\begin{abstract} 
Exact calculations of the transmittance  of 
surface corrugated optical waveguides are presented. 
The elastic scattering of diffuse light or other electromagnetic waves from a 
 rough surface induces a diffusive transport along 
the waveguide axis. As the length of the corrugated part of the waveguide 
increases, a transition from the {\em diffusive} to the {\em localized} 
 regime is observed. This involves an analogy with electron
conduction in nanowires, and hence, a concept  analogous to that of
 ``resistance'' can be introduced. We  show 
an oscillatory behavior of both the elastic mean free path and the  
localization length versus the wavelength. \\
\end{abstract}
] 

The influence of inhomogeneities and 
defects in the transmission properties of waveguides for both quantum 
and classical waves is a subject of increasing 
research\cite{Beenaker,beenaker2,MicNat91,TamAnd,McKPRB93,NieEPL}. 
The consequences of volume defects on electron wave propagation through
two dimensional (2D) wires have been 
recently 
addressed\cite{PelosPRB96,NiePRB96,TodPRB96}. On the other hand, 
some aspects of the effects of slight surface roughness and volume 
inhomogeneities both in optical fibers and waveguides have been under 
study for a long time\cite{Marcuse,Hall}, mainly in connection with their 
attenuation;  however, the study of these effects has been to date 
concentrated on the wave scattering into the radiation field; 
therefore, no analysis has been made so far of the coupling between 
forward and backward modes\cite{Ladou}, neither on the influence of 
surface roughness induced localization effects\cite{McGurn} into 
both this coupling and the radiation losses. 

Although some analogies
between electron waves and light and other classical waves have been 
put forward in apertures\cite{beenaker2}, in this letter we 
address   localization effects of light and other electromagnetic waves 
propagating through a 2D corrugated waveguide (see inset of Fig. 1). 
There are two main differences between
electron transport and the common situation in optical guides. 
First, the electron  waves  are confined within the conductor and, 
for small voltages, they {\em do not couple} with the radiation field. 
Second, the electron modes are mutually {\em incoherent}. 
So to resemble this picture we consider a waveguide with {\em perfectly 		
conducting} walls, and the incident modes 
to be mutually {\it incoherent}. This can be achieved for example by 
previously passing the incident
light through a moving diffuser so that correlation between phases of different
modes is lost. 

The corrugated part of the waveguide, of total length $L$, is   composed 
of $n$ slices of length $l$. 
The width of each slice has random values uniformly distributed between 
$W_0-\delta$ and $W_0+\delta$ about a mean value $W_0$.  We shall
 take $W_0/\delta= 7$ and $l/\delta = 3/2$ (i.e. 
$2\delta/W_0 = 0.286$ and $l/W_0 = 0.214$). The main transport properties 
do not depend on the particular choice of these parameters, however. 
s-polarization
 with the electric vector parallel to the grooves (TE modes) is assumed. 
Transmission and reflection coefficients are exactly calculated 
on solving the 2D wave equation by mode matching at each slice, together 
with a generalized scattering-matrix technique, whose details can be 
found in Ref. 14.

We define the normalized transmittance of the waveguide  as 
$G = \sum_{ij} T_{ij}$, 
where $T_{ij}$ is the ratio between the total   flux transmitted from 
the incoming mode $i$ 
into the outgoing mode $j$, $\Phi_{j}^{out}$, and  the total flux of 
this incoming mode, 
$\Phi_{i}^{in}$, i.e. \mbox{ $T_{ij} = \Phi_{j}^{out}/\Phi_{i}^{in}$,} 
and the sum runs over the total number of propagating modes. 
We also  define the optical analogue of the ``resistance'' as the inverse 
of the transmittance\cite{NiePRB96} $R = 1/G$. 
Ensemble averages, denoted by $\langle . \rangle$, are performed over 100
realizations of the corrugated waveguide. 

For a perfect waveguide of constant width $W_0$, the normalized 
transmittance $\langle G \rangle$ is simply given by the number of propagating 
modes $N$, 
i.e. 
$\langle G \rangle=N$ = E$\{2 W_0/\lambda\}$. A plot of $\langle G \rangle$ 
versus $W_0/\lambda$ (see 
Fig.1) shows a
 staircase behavior which constitutes the analogous for classical waves of the 
conductance quantization of electronic systems \cite{Nature}. 
Fig. 1  also contains $\langle G \rangle$ versus $W_0/\lambda$ 
for different length values $L$ of the corrugated portion of the 
waveguide. 
For moderate lengths $L$, the transmittance shows a dip just at the onset of 
a new propagating mode. 
Similar dips appear in the case of scattering 
by volume defects\cite{NieEPL,dips}. 

As the length $L$ of the 
corrugated region increases, 
$\langle G \rangle$ presents an oscillatory behavior that, as we show below, 
reflects successive transitions from {\it diffusive} to {\it localized} 
regimes. This is done on analysing
the dependence of $\langle R \rangle$ and 
$\langle \ln (G)  \rangle$ on  $L$. Note that, while in clean 
systems 
the transport is {\em ballistic} and consists essentially of  unscattered 
waves, (the mean free path $\ell$ is then much larger than both $L$ and 
$W_0$), in a randomly corrugated waveguide the two regimes, aforementioned above, are clearly visible as shown in 
Fig. 2, which contains $\langle R \rangle$ and 
$\langle \ln (G)  \rangle$ versus  $L$, (cf. Fig. 2(a) and 2(b), 
respectively), for $W_0/\lambda =  2.6$. As $L$ increases from zero,
$\langle R\rangle$ follows first an 
almost perfect linear behavior with $L$ which corresponds to the 
{\em semi-ballistic} \cite{NieEPL} regime.
In this case, there is so much 
scattering in the longitudinal direction that the transport is 
{\em diffusive} while in the transversal direction almost no scattering occurs 
($W_0 < \ell < L$). 
Then, in analogy with Ohm's law for electron wires, the 
 waveguide ``resistance'' can be described by an 
``ohm-like'' term proportional to $L/W_0$ plus a ``contact 
resistance'' $R_c$\cite{NiePRB96,TodPRB96}, 
\begin{equation} 
\langle R \rangle = R_c + \rho \frac{L}{W_0} = R_c + \frac{L}{N\ell}~~,
\label{conduc} 
\end{equation} 
where the ``resistivity'', $\rho$, and the mean 
free path $\ell$ are related by\cite{TodPRB96} $\ell/W_0 =  1/(N\rho)$.

This linear behavior breaks down as $L$ increases further. 
 Then, a typical linear decrease of 
$\langle \ln (G)  \rangle$ with $L$ appears, as shown in 
Fig. 2(b). The system has now entered in the {\it localized 
regime} at which $W_0 < \ell \ll L$.
 The localization length 
$\xi$ is defined by\cite{Localization,Lee} 
\begin{equation} 
\xi \equiv - \left(\frac{\partial \langle \ln (G)  \rangle}{\partial 
L}\right)^{-1}~~.
\end{equation} 

The contact resistance 
$R_c$ and the effective mean free path $\ell/W_0$ 
can be easily obtained both
from Eq. \ref{conduc} and by  a least-square fitting of the linear 
part in Fig. 2(a). 
In  Fig. 3(a) we show these values of $R_c$ and $\ell/W_0$ versus 
$W_0/\lambda$. 
The contact resistance (see the inset of Fig. 3) is slightly higher than 
the expected value $1/N$, due to the transition region between ballistic 
and semi-ballistic regimes. 
The mean free path oscillates 
as $W_0/\lambda$ increases,   having its minima close to half-integers 
of $W_0/\lambda$, which correspond to the appearence of a new 
propagating mode in the waveguide, and thus establishes the above quoted 
connection with the transmittance dips at the onset of new modes 
shown in Fig. 1. 

The dependence of the localization length $\xi$ on the 
wavelength (Fig. 3(b)) can be also obtained 
from   least-square fitting of the linear part of the plot of 
$\langle \ln (G)  \rangle$ versus $L$ in Fig. 2(b). 
It should be noted that, 
within 
the numerical accuracy of our calculation, 
$\xi \approx N\ell$ 
as  shown in Fig. 3(b). This again has an analogy 
in agreement with the expected behavior for electron 
transport through disordered media 
\cite{Thouless,Pichard}. 
 
The wavelength dependence of both the localization length and the mean free
path shows that, for a  
fixed length, it is possible to pass on from the localized regime into the 
diffusive one by  changing the wavelength of the incoming wave. 

In conclusion, we have shown that diffuse light and other electromagnetic wave 
propagation through surface corrugated perfectly reflecting waveguides, 
constitutes the optical analogue of electron transport in disordered nanowires.
As such, the concept of ``averaged resistance'' and the optical analogous of
 ``Ohmic'' behavior in the 
semi-ballistic (diffusive) regime can be introduced for these systems. 
We have also shown 
that both the effective mean free path and the localization length 
oscillate with the wavelength. 
This oscillatory transition from diffusive to 
localized behavior should be observed in optical waveguides. 
The influence of surface induced localizaton effects on radiation losses  in actual waveguides 
remains an open question. We hope 
that our results will estimulate both experimental and theoretical research in this direction.

We would like to thank A. Caama\~{n}o, E. Dobado-Fuentes, T. L\'{o}pez, 
Th.M. Nieuwenhuizen and F. Sols for interesting 
discussions.  This work has been supported by the Direcci\'{o}n General 
Cient\'{\i}fica y T\'{e}cnica (DGICyT) through Grant No.  PB95-0061. A. G-.M. aknowledge
partial financial support from the postgraduate grant program of the Universidad Aut\'{o}noma
de Madrid.

\begin{figure} 
\caption{ 
Averaged transmitance  $\langle G \rangle$ as a 
function of $W_0/\lambda$ for  fixed values of 
$\delta/W_0$ and $l/W_0$, 
($W_0=7\, \delta$ and $l=(3/2)\, \delta$). Note that the case $L/W_0=0$
corresponds to a flat waveguide. 
The inset shows a 
schematic view of the system. 
} 
\label{fig1} 
\end{figure} 

\begin{figure} 
\caption{ 
(a) $\langle R\rangle$ versus the length $L$ of the corrugated region. 
The straight line represents the best fitting of the linear behavior 
associated to the {\em diffusive} transport regime (see Eq.\ref{conduc}). 
The results are obtained for 
$W_0/\lambda = 2.6$, i.e they correspond to 5 propagating modes in the 
waveguide. The corresponding 
effective mean free path is $\ell/W_0 
\approx 
6.3 $.
(b)  $\langle 
\ln (G) \rangle$ versus $L$ for the 
same case as in (a). The best 
fitting of the linear part of the curve is also shown. The 
corresponding localization length is $\xi/W_0 \approx 34.4$. 
} 
\label{fig2} 
\end{figure} 

\begin{figure} 
\caption{ 
(a) 
Effective mean free path $\ell/W_0$ versus $W_0/\lambda$. 
The inset shows the behavior of the contact resistance 
$R_c$ together with the expected value for a 
perfect waveguide (dotted line). 
(b) 
Localization length $\xi/W_0$ (filled circles) and $N\ell$ (open 
circles) versus $W_0/\lambda$.  
$\xi \approx N\ell$, within the numerical accuracy. 
} 
\label{fig3} 
\end{figure} 
\end{document}